\documentclass[twocolumn,english,aps,prb,floats]{revtex4}
\usepackage[T1]{fontenc}
\usepackage[latin9]{inputenc}
\usepackage{bm}
\usepackage{amsmath}
\usepackage{amssymb}
\usepackage{graphicx}
\usepackage{esint}
\usepackage{adjustbox}
\usepackage{longtable}
\makeatletter
\@ifundefined{textcolor}{}
{%
 \definecolor{BLACK}{gray}{0}
 \definecolor{WHITE}{gray}{1}
 \definecolor{RED}{rgb}{1,0,0}
 \definecolor{GREEN}{rgb}{0,1,0}
 \definecolor{BLUE}{rgb}{0,0,1}
 \definecolor{CYAN}{cmyk}{1,0,0,0}
 \definecolor{MAGENTA}{cmyk}{0,1,0,0}
 \definecolor{YELLOW}{cmyk}{0,0,1,0}
}

\usepackage{babel}

\makeatother

\usepackage{babel}
\begin{document}

\title{Scaling Theory of Magnetic Order and Microwave Absorption in Amorphous and Granular Ferromagnets}
\author{Eugene M. Chudnovsky and Dmitry A. Garanin}
\affiliation{Physics Department, Herbert H. Lehman College and Graduate School,
The City University of New York, 250 Bedford Park Boulevard West,
Bronx, New York 10468-1589, USA }
\date{\today}
\begin{abstract}
Magnetic order and microwave absorption in amorphous ferromagnets and materials sintered from nanoscale ferromagnetic grains are investigated analytically and numerically within the random-anisotropy model. We show that a scaling argument specific to static randomness allows one to make conclusions about the behavior of a large system with a weak disorder by studying a smaller system with a strong disorder. The breakdown of the scaling on increasing the strength of the magnetic anisotropy and/or the size of the grain separates two distinct regimes in magnetic ordering and frequency dependence of the absorbed microwave power. Analytical results are confirmed by numerical experiments on spin lattices containing up to ten million spins. Our findings should help design materials with desired magnetic and microwave properties. The method can be extended to other systems with quenched randomness.
\end{abstract}
\maketitle

\section{Introduction}
\label{Intro}

Static randomness is a feature of many systems with a continuous order parameter. They include amorphous and sintered magnets  (see, e.g., Refs.\ \onlinecite{RA-book,CT-book,Marin-MMM2020,GC-JPhys2022} and references therein), arrays of magnetic bubbles \cite{bubbles}, pinned charge-density waves \cite{Efetov-77,Okamoto-2015} and flux lattices in superconductors \cite{Blatter-RMP1994,EC-PRB1989}, liquid crystals and polymers \cite{LC}, thin films on imperfect substrates \cite{EC-PRB1986}, superfluid $^3$He-A in aerogel \cite{Volovik-JLTP2008,Li-Nature2013,Volovik-JETPlett2018}, and others. The effect of static disorder on the order parameter is stronger than the effect of thermal fluctuations \cite{IM}. It usually destroys the long-range order even at zero temperature on a scale that is inversely proportional to the disorder's strength. Consequently, observation of the effect of weak static randomness on the long-range order requires a large system. This often inhibits numerical studies of the effect because they demand a prohibitively large computing time even with the use of the most powerful computers. In this paper, using an example of amorphous and granular ferromagnets described by the random-anisotropy (RA) model \cite{Harris-PRL1973}, we demonstrate how this problem can be solved by applying scaling that is specific to static randomness. Both statics and dynamics of random magnets will be studied and practical results based on that scaling method will be derived that must assist experimentalists in designing materials with desired magnetic and microwave properties. 

Theoretical research on amorphous and sintered magnets has been largely based on the Imry-Ma concept \cite{IM}. Amorphous ferromagnets are obtained by rapid quenching from the melt of a substance that would become a crystalline ferromagnet on slow cooling. This does not allow the neighboring atoms to develop a crystalline lattice. Nevertheless, the neighboring spins still develop parallel orientation due to the strong overall ferromagnetic exchange. Directions of local magnetic anisotropy axes are random, however, or are correlated at short distances determined by the amorphous structure factor. The RA strength $D_R$ is typically small compared to the exchange $J$ because magnetic anisotropy is produced by the relativistic spin-orbit interaction while the exchange is due to the Coulomb interaction between electrons \cite{CT-book}.  Nevertheless, as suggested by Imry and Ma, weak random pushes from the magnetic anisotropy make the direction of the magnetization slightly wander from one spin to the other, disordering it at large distances as in the random walk problem. The corresponding ferromagnetic correlation length is $R_f \sim (J/D_R)^{2/(4-d)}a$, with $a$ being the average interatomic distance and $d$ being the dimensionality of the system, implying the destruction of the long-range order in less than four dimensions. This magnetic state received the name of the correlated spin glass (CSG) \cite{CSS-1986,CT-book,PCG-2015}. 

The question of the ground state of the RA model has never been settled to everyone's satisfaction. Numerous studies of systems with static disorder \cite{Cardy-PRB1982,Villain-ZPB1984,Nattermann,Kierfield,Korshunov-PRB1993,Giamarchi-95,LeDoussal-PRL07,Bogner,Orland-EPL,Garel-PRB} that used renormalization-group, scaling, replica-symmetry-breaking, and variational methods often suggested types of the ordering that deviated from the random-walk picture. So did numerical studies \cite{Gingras-Huse-PRB1996,Zeng,Rieger,Itakura-05,Perret-PRL2012,Fisch-1998,Fisch-2000} that assumed full thermal equilibrium. This effort subsided after it was realized that from the practical perspective, the problem of the ground state (or the low-temperature state) of a metastable system with quenched randomness and an exponentially large number of local energy minima was largely irrelevant. Besides, even if the full thermal equilibrium was achieved, the long-range correlations would be strongly affected by the presence of topological defects \cite{GCP-PRB2013,PGC-PRL,CG-PRL}. The low-temperature state of any many-body system that exhibits hysteresis depends on history and initial condition. 

Ferromagnets, and RA ferromagnets in particular, are a good example of that \cite{Serota-1986,Dieny-PRB1990,DC-PRB1991,GC-JPhys2022}. Since the uniaxial anisotropy, unlike the magnetic field, has two preferred directions, one can argue that the global ordering of all spins into a hemisphere should be preferred energetically in the RA magnet due to the ferromagnetic exchange. In real experiments, as well as in the numerical work on systems below four dimensions, however, the magnetic state of an RA ferromagnet  depends on the initial condition. Starting with a collinear initial condition (CIC), that is with all spins looking in the same direction, the minimization of the energy in zero field leads to the state with a finite magnetization. On the contrary, energy minimization from a random initial condition (RIC), with spins oriented randomly (which mimics cooling down from a paramagnetic state), leads to a fully disordered state with the magnetization wandering on a sphere on the spatial scale $R_f$ that decreases on increasing $D_R/J$. For a magnetized state, that evolved from the CIC in the absence of the field, $R_f$ describes the scale of wandering of the transversal component of the magnetization. 

Until now, computer simulations of the RA problem, limited by the system size, addressed values of $D_R/J$ that were large compared to $D_R/J \sim 10^{-6} - 10^{-3}$ relevant to disorder at the atomic scale in amorphous magnets. In this paper, we will show that due to the scaling, the physical case of a very small $D_R$ is mathematically equivalent to $D_R \sim J$. This  allows one to greatly reduce the computing time when studying the RA model numerically in application to amorphous ferromagnets. A similar advantage of the proposed scaling must exist for many other systems with the static disorder. For a granular ferromagnet, the effective $D_R$ can be large. We shall demonstrate that there exists a critical value of $D_R/J$, that is, a critical grain size, at which the scaling breaks down and the behavior of the RA system changes qualitatively. 

The above picture relates to the static properties of random magnets that have been intensively investigated in the past. By comparison, studies of their dynamic properties were scarce. They have been mainly focused on the ferromagnetic resonance \cite{Saslow2018} (FMR) for which there are experimental data \cite{Monod,Prejean,Alloul1980,Schultz,Gullikson}, and on the the localization of spin modes \cite{Fert,Levy,Henley1982,HS-1977,Saslow1982,Bruinsma1986,Serota1988,Ma-PRB1986,Zhang-PRB1993,Alvarez-PRL2013,Yu-AnnPhys2013,Nowak2015} that has been reported in various magnetic systems with static disorder \cite{Amaral-1993,Suran1-1997,Suran2-1997,Suran-1998,McMichael-PRL2003,Loubens-PRL2007,Du-PRB2014}.  

More recently, it was shown \cite{GC-PRB2021,GC-PRB2022,GC-PRB2023L,CG-PRB2023IP}  that nonconducting amorphous ferromagnets, or granular materials comprised of coated magnetic particles of size below the skin depth, can be promising systems for strong broadband microwave absorption. In this paper, we apply the scaling method to make predictions about the dependence of the absorbed microwave power on frequency, RA, and the size of the grain. We test these predictions in numerical experiments using the model of atomic spins on the lattice with RA and obtain good agreement with the scaling theory The numerical method we used is described in detail in Refs.\ \onlinecite{GC-PRB2021,GC-PRB2022}.

The paper is organized as follows. Scaling of the static properties of the RA magnets is discussed in Section \ref{RA-magnet}. It is worked out for an amorphous ferromagnet with the atomic-scale disorder in Subsection \ref{amorphous}), and for a magnet sintered of nanoscale ferromagnetic grains in Subsection \ref{granular}. Evidence of the transition on the strength of the RA  and/or the size of the grain is presented. Microwave absorption by granular ferromagnets is studied in Section \ref{absorption}, with the scaling argument for the peak frequency presented in Subsection \ref{peak} and the scaling of the power absorption worked out in Subsection \ref{power}.  Our findings are summarized in Section \ref{discussion}. 

\section{Magnetic order in a random-anisotropy ferromagnet}
\label{RA-magnet}
\begin{figure}[h]
\centering{}\includegraphics[width=8cm]{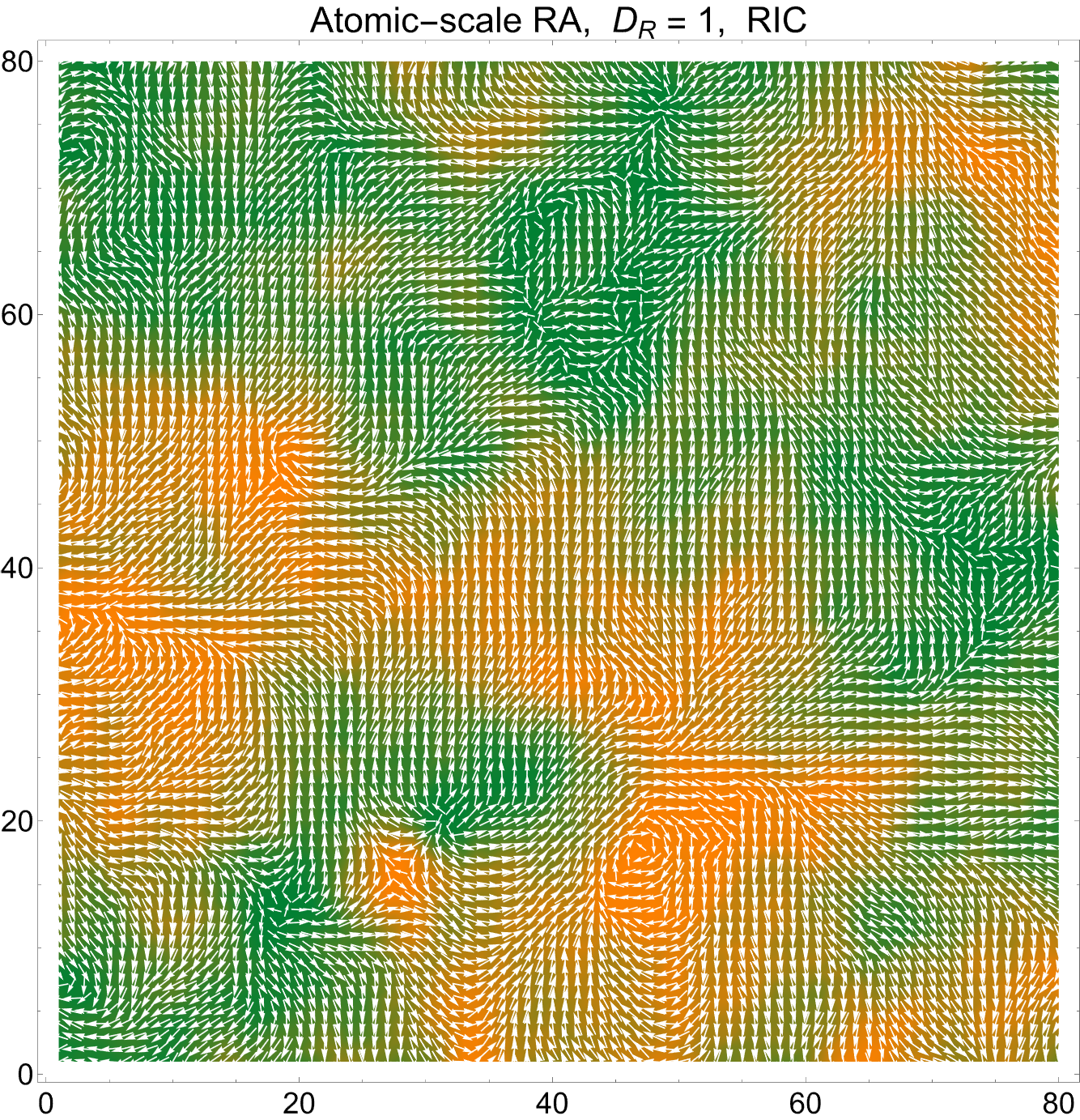}\
\caption{Equilibrium spin structure obtained numerically in a 2D amorphous ferromagnet with random anisotropy axes of individual spins and RIC. In-plane spin components are shown by white arrows. The out-of-plane component is shown by orange/green corresponding to positive/negative.}
\label{fig-amorphous} 
\end{figure}

The RA model reflects the fact that the exchange interaction and the magnetic anisotropy are determined by the local atomic environment. If the exchange is predominantly ferromagnetic (favoring parallel orientation of neighboring spins), in the first approximation one can choose a single constant $J$ for all pairs of neighboring spins.  Similarly, one can choose a single constant $D_R$ for the strength of the magnetic anisotropy for all spins. However, its direction ${\bf n}$ must be random due to the absence of the global crystallographic order. The profound effect on the magnetic ordering comes from that random orientation of the anisotropy axes. In what follows we shall distinguish between an RA magnet with anisotropy axes disordered at the scale of individual spins, see Fig.\ \ref{fig-amorphous}, and the magnet sintered from randomly oriented nanocrystallites. We begin with the first case. 

\subsection{Amorphous ferromagnet disordered at the atomic scale}
\label{amorphous}
Here we consider an amorphous ferromagnet described by the RA model in $d$ dimensions with the Hamiltonian
\begin{equation}
{\cal{H} } = \frac{1}{2}\int \frac{d^d r}{a^d}\left[Ja^2(\nabla {\bf s})^2 - D_R({\bf n} \cdot {\bf s})^2\right],
\label{Hamiltonian}
\end{equation}
where ${\bf s}({\bf r})$ is the unit vector of the dimensionless spin-field density and ${\bf n}({\bf r})$ is the unit vector of the magnetic anisotropy that is assumed to be random on a spatial scale $a$ representing the average interatomic distance. 

The Imry-Ma argument describing the effect of this type of disorder on magnetic ordering goes like this. Let $R_f$ be the characteristic distance where the directions of spins begin to deviate strongly from the ferromagnetic alignment. From the Hamiltonian (\ref{Hamiltonian}) the density of the exchange energy per spin would be of order $J\left({a}/{R_f}\right)^2$, and the total exchange energy in a system of size $L$ in $d$ dimensions would be 
\begin{equation}
E_{ex}\sim J\left(\frac{a}{R_f}\right)^2\left(\frac{L}{a}\right)^d.
\end{equation}
If directions of the anisotropy axes were uniformly distributed on a $d$-dimencional  sphere, the anisotropy energy per spin would be a constant $D_R/d$  due to $\langle n_i n_j \rangle = \delta_{ij}/d$. The dependence of the anisotropy energy on $R_f$ comes from statistical fluctuations in the distribution of anisotropy axes which define the direction of average magnetization in the volume of size $R_f$. They add the term of order $E_{an} \sim -D_R\left({a}/{R_f}\right)^{d/2}$ to the anisotropy energy per spin, which, in turn, adds
\begin{equation}
E_{an} \sim -D_R\left(\frac{a}{R_f}\right)^{d/2}\left(\frac{L}{a}\right)^d
\end{equation}
to the total anisotropy energy. The energy minimum of $E = E_{ex} + E_{an}$ on $R_f$ is achieved at
\begin{equation}
R_f = k_d \left(\frac{J}{D_R}\right)^{{2}/{(4-d)}} a,
\label{Rf}
\end{equation}
where $k_d$ is a $d$-dependent numerical factor. The ferromagnetic correlation length $R_f$ is finite for all $d < 4$, no matter how weak the RA is. Various approximations show that $k_d$ is large compared to one so that the condition of validity of Eq.\ (\ref{Rf}), $R_f \gtrsim a$, is satisfied even for $D_R > J$. 
A rough estimate of the limiting value of $D_R$  can be obtained from the condition that the anisotropy energy of an individual spin,  $D_R$, is less than its total exchange energy $2dJ$ with the nearest neighbors.  Above that value of the RA, the problem simplifies: directions of spins are determined by the directions of the local anisotropy axis. 

\begin{figure}[h]
\centering{}\includegraphics[width=9cm]{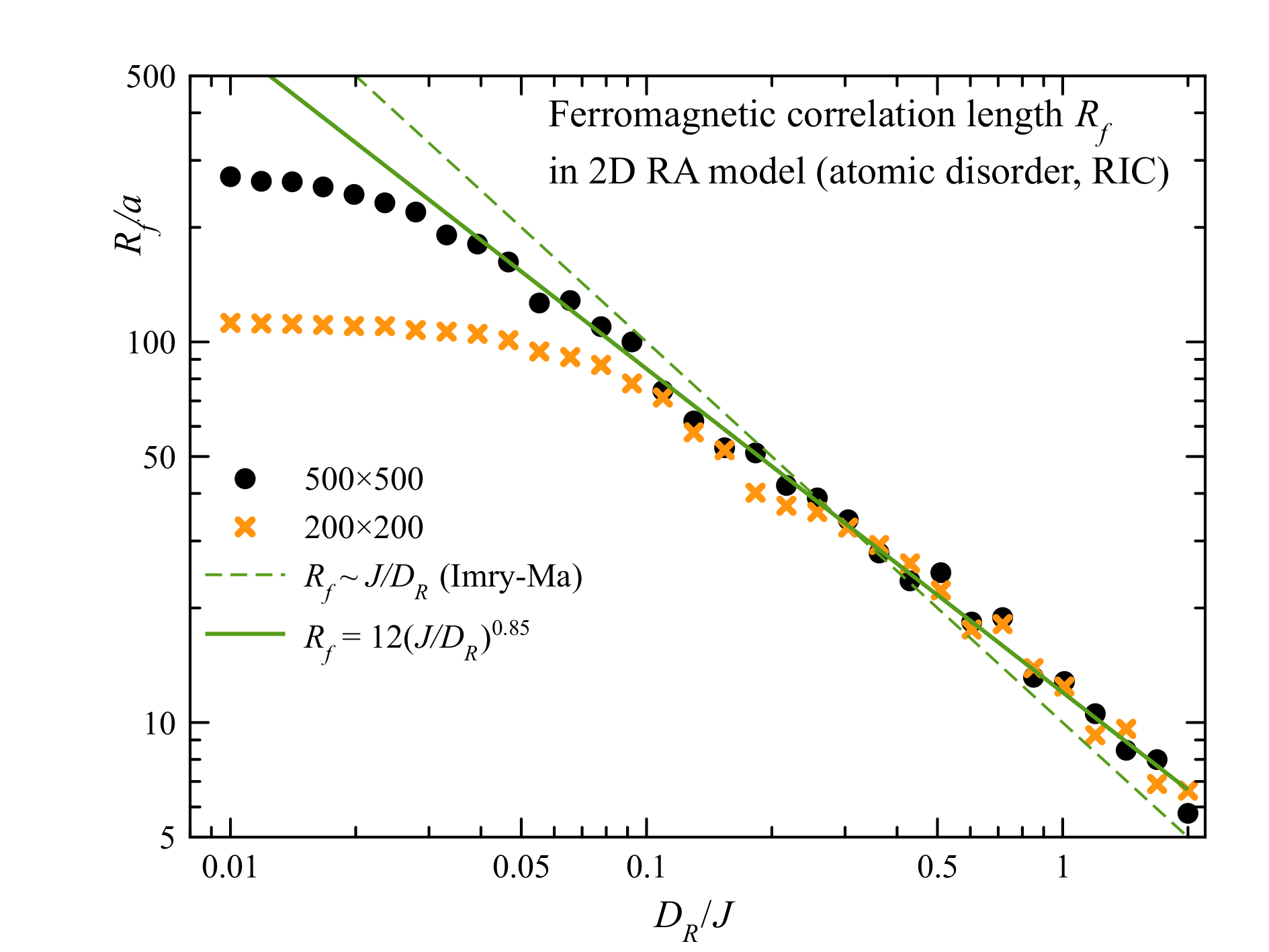}\
\caption{Dependence of the ferromagnetic correlation length on the strength of random anisotropy in two dimensions.}
\label{Rf-vs-DR} 
\end{figure}

The dependence of $R_f$ on $D_R/J$ in two dimensions (thin film), computed on lattices of size $200 \times 200$ and $500 \times 500$ is shown in Fig.\  \ref{Rf-vs-DR}. It has been obtained from the general formula for the fluctuations of the magnetization \cite{GC-JPhys2022}, ${\bf m} = N^{-1}\sum_{i}{\bf s}_i$, in a finite-size system of $N$ spins, 
\begin{equation}
\langle m^2\rangle = \frac{1}{N}\sum_{j}\langle{\bf s}_i \cdot {\bf s}_{i+j}\rangle = \frac{1}{N}\int_0^\infty \frac{d^d r}{a^d} C(r).
\end{equation}
For a Gaussian spin-spin correlation function $C(r) = \exp(-r^2/R_f^2)$ that we used it yields $m^2 = \pi R_f^2/(Na^2)$. The power $-0.85$, obtained numerically  and shown in Fig.\ \ref{Rf-vs-DR}, differs from power $-1$ predicted by theory. We attribute it to the energy barriers and topological defects seen in Fig.\ \ref{fig-amorphous} that are not accounted for in the Imry-Ma approximation. They prevent the system from relaxing to the IM state  \cite{PGC-PRL}. The difference in the power is not large, which allows one to use scaling formulas based on the IM concept with fair accuracy. Deviation observed on decreasing $D_R/J$, which depends on the system size, is clearly related to the violation of condition $R_f \ll L$ used for deriving theoretical formulas.

A remarkable property of the RA model is that it is non-perturbative on $D_R/J$. As we show below, the case of $D_R \ll J$ is mathematically equivalent to $D_R \sim J$. Thus, the RA problem is a strongly correlated problem regardless of the RA strength. The only difference is in the ferromagnetic correlation length $R_f$. On one hand, this makes it difficult to compute the exact spin-spin correlation function analytically. On the other hand, it greatly simplifies the numerical work. Indeed, for $D_R \ll J$ of practical interest, $R_f$ can be very large. To distinguish such a state from the long-range ferromagnetic order, one needs to do computations on a system of size $L \gg R_f$ containing a very large number of spins. Luckily, the mathematical equivalence of this case to $D_R \sim J$ allows one to have an insight into the properties of the system by considering a much smaller $R_f$. With that purpose, instead of considering unit cells of size $a$ with the local exchange $J$ and the local anisotropy $D_R$, let us consider blocks of spins of size $r$ satisfying $a<r<R_f$, coupled  by the exchange $J_{eff}$ and characterized by the effective anisotropy $D_{eff}$. The corresponding rescaled exchange and anisotropy energy become
\begin{equation}
E'_{ex} \sim J_{eff}\left(\frac{r}{R_f}\right)^2\left(\frac{L}{r}\right)^d\end{equation}
and
\begin{equation}
E'_{an} \sim -D_{eff}\left(\frac{r}{R_f}\right)^{d/2}\left(\frac{L}{r}\right)^d
\end{equation}
Equating the expressions for exchange and anisotropy energies before and after rescaling, we obtain
\begin{equation}
J_{eff} = J \left(\frac{r}{a}\right)^{d-2}, \qquad D_{eff} = D_R\left(\frac{r}{a}\right)^{d/2}.
\label{eff}
\end{equation}
The last equation for $D_{eff}$ is an obvious consequence of statistical fluctuations of anisotropy directions inside the block of spins of size $r$. The expression for $R_f$ that follows from the minimization of   $E' = E'_{ex} + E'_{an}$ is the same as before
\begin{equation}
R_f = k_d\left(\frac{J_{eff}}{D_{eff}}\right)^{{2}/{(4-d)}} r = k_d\left(\frac{J}{D_R}\right)^{{2}/{(4-d)}} a,
\end{equation}
confirming the invariance of the problem with respect to rescaling to bigger blocks. It must be related to the fractal structure of the random walk that is behind the Imry-Ma argument. 

Since the scaling requires $a < r < R_f$, it is clear that it breaks at $R_f \sim a$, that is, at 
\begin{equation}
D_R = D_R^{(c)} \sim k_d^{(4-d)/{2}} J
\end{equation}
on increasing $D_R$. The practical consequence of this is that for any $D < D_c$, the answer for any global characteristic of the RA system can be obtained numerically by making computation at $D = D_c$, which should provide an enormous computational advantage because at $D = D_c$ the condition $R_f < L$ can be satisfied for a relatively small system as compared to a very large system needed to satisfy that condition at $D \ll J$. 

\begin{figure}[h]
\centering{}\includegraphics[width=9cm]{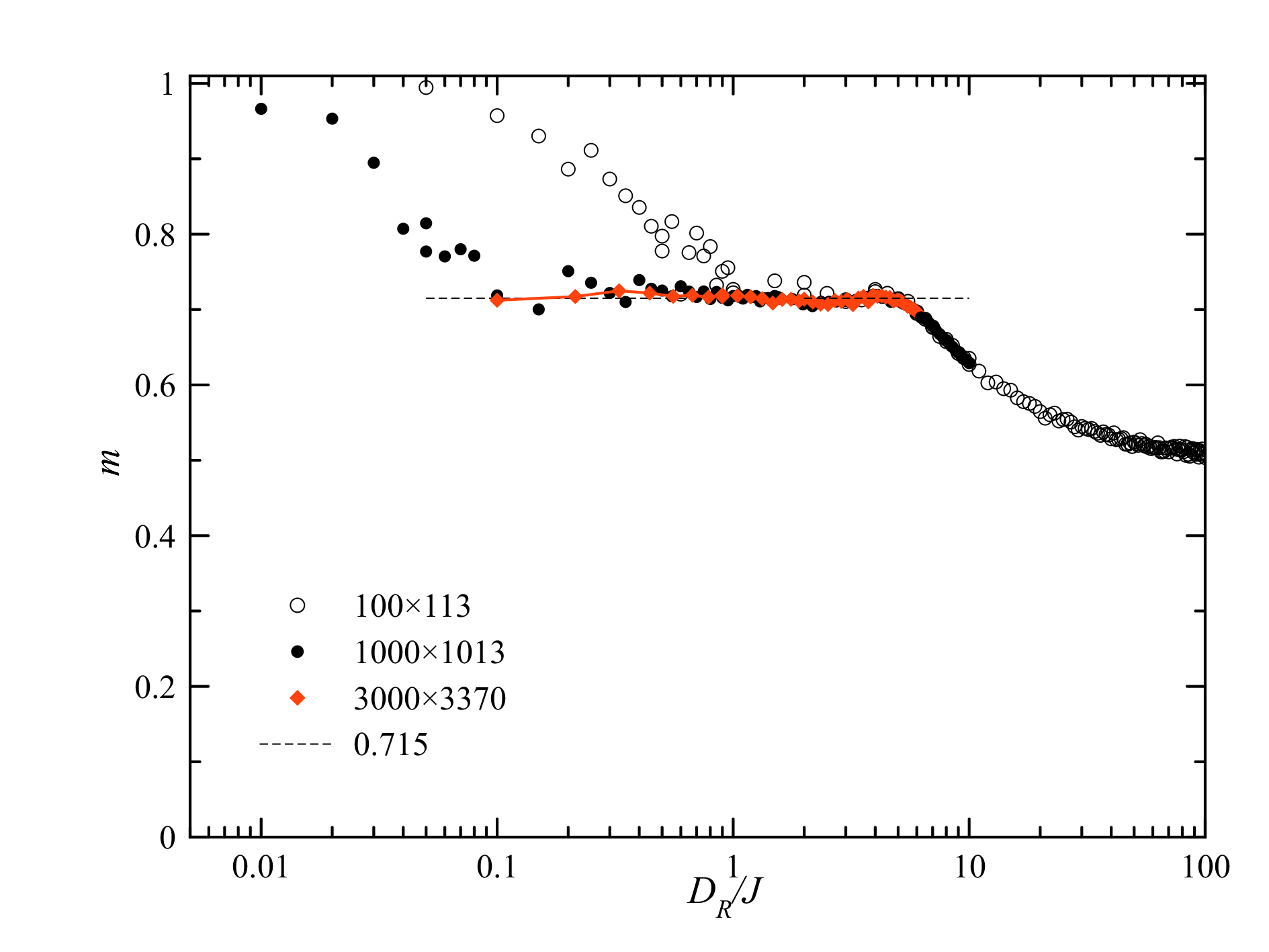}\
\caption{Magnetic moment per spin in a two-dimensional RA model, obtained numerically by relaxation from the initial state with collinear spins.}
\label{transition} 
\end{figure}

Fig.\ \ref{transition} shows the average magnetic moment per spin of a two-dimensional RA system, obtained by relaxing from the initial state with collinear spins. This can be achieved by aligning all spins in a strong magnetic field and then switching the field off. As soon as the condition $R_f \ll L$ is satisfied, the curves in a broad range of $D_R$ tend to the universal value of $m = 0.715$ per spin, confirming the scaling property of the RA model demonstrated above. It breaks at the critical value of the RA that is close to $D_R^{(c)} = 5J$. Substituting it into the fit $R_f/a = 12(J/D_R)^{0.85}$ shown in Fig.\ \ref{Rf-vs-DR}, we get $R_f \approx 3a$ for the value of the ferromagnetic correlation length below which the scaling breaks down. On increasing $D_R$ the magnetic moment per spin tends to $m = 0.5$ as is expected for three-component spins directed randomly in a hemisphere.

Similar formulas can be easily generalized for the RA film of finite thickness $h$. Here the exchange energy per spin remains the same, $J({a}/{R_f})^2$,
but the anisotropy energy becomes modified by statistical fluctuations in the direction normal to the film. This adds the factor $(a/h)^{1/2}$ to the formula (\ref{eff}) for $D_{eff}$ with $d = 2$, making the anisotropy energy per spin of the film 
$-D_R({a}/{h})^{1/2}({a}/{R_f})$. The minimum of the total energy is achieved at
\begin{equation}
\frac{R_f}{a} \sim \left(\frac{h}{a}\right)^{1/2}\frac{J}{D_R}.
\end{equation}
This formula is valid for $h < R_f$. At $h = R_f$ it gives
\begin{equation}
\frac{R_f}{a} \sim \left(\frac{J}{D_R}\right)^2,
\end{equation}
as it should be in three dimensions.

\subsection{Granular ferromagnet}
\label{granular}

\begin{figure}[h]
\centering{}\includegraphics[width=8cm]{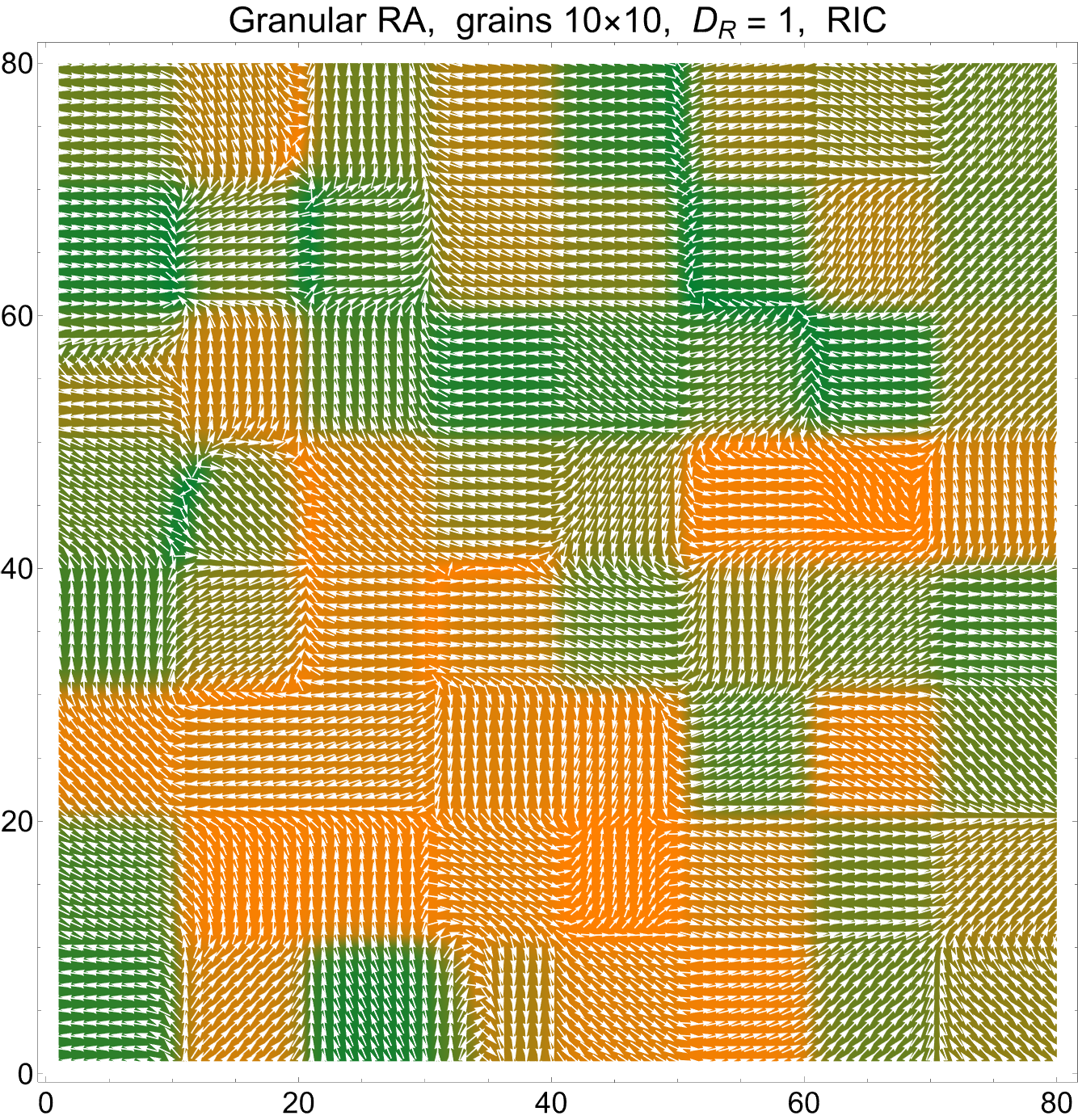}\
\caption{Spin structure of a ferromagnet sintered from randomly oriented nanograins having the same anisotropy direction for all spins inside the grain. The RIC has been used. The color code is the same as in Fig.\ \ref{fig-amorphous}.}
\label{grains} 
\end{figure}
The scaling argument that is similar to the one presented in the previous subsection can be developed for a granular ferromagnet sintered from nanocrystals. In practice, this problem introduces another source of randomness due to the random shape and size of the nanocrystals. Here we consider a model of a granular ferromagnet made of identical densely packed nanocrystals (see Fig.\ \ref{grains}) with randomly oriented easy anisotropy axes. We believe that such a simplified model correctly catches static and dynamic properties of sintered ferromagnets. Due to the same anisotropy strength and direction for all spins inside the individual grain, it permits rigorous rescaling from atomic spins to the spins of the grains that can be tested numerically. 

The argument goes like this. Keeping the spin density ${\bf s}^2 =1$ for the grains, Hamiltonian (\ref{Hamiltonian}) can be rescaled by writing it in terms of the rescaled exchange $J'$ and anisotropy $D'_R$ acting on the spins of the grains of size $R_a$,
\begin{equation}
{\cal{H}} = \frac{1}{2}\int \frac{d^d r}{R_a^d}\left[J'R_a^2(\nabla {\bf s})^2 - D'_R({\bf n} \cdot {\bf s})^2\right] .
\label{Hamiltonian-Ra}
\end{equation}
Hamiltonians (\ref{Hamiltonian}) and (\ref{Hamiltonian-Ra}) can be understood as continuous limits of the corresponding discrete problems with atomic and grain spins. Comparing Eq.\ (\ref{Hamiltonian-Ra}) with the original expression (\ref{Hamiltonian}) for $R_a = a$, we obtain
\begin{equation}
J' = J\left(\frac{R_a}{a}\right)^{d-2}, \qquad D'_R = D_R\left(\frac{R_a}{a}\right)^d
\label{scaling}
\end{equation}
For the atomic disorder defined by $R_a = a$, the ferromagnetic correlation length is given by Eq. (\ref{Rf}). The same argument must apply to the blocks of spins of size $R_a$. This gives
\begin{equation}
\frac{R_f}{R_a} = k_d \left(\frac{J'}{D'_R}\right)^{2/(4-d)} = k_d \left(\frac{J}{D_R}\right)^{2/(4-d)}\left(\frac{a}{R_a}\right)^{4/(4-d)},
\label{Rf-Ra}
\end{equation} 
\begin{equation}
\frac{R_f}{a} = k_d \left(\frac{J}{D_R}\right)^{2/(4-d)}\left(\frac{a}{R_a}\right)^{d/(4-d)}.
\label{Ra}
\end{equation}
The ferromagnetic correlation length goes down as  $1/R_a$ in two dimensions, and as $1/R_a^3$ in three dimensions. 

\begin{figure}[h]
\centering{}\includegraphics[width=9cm]{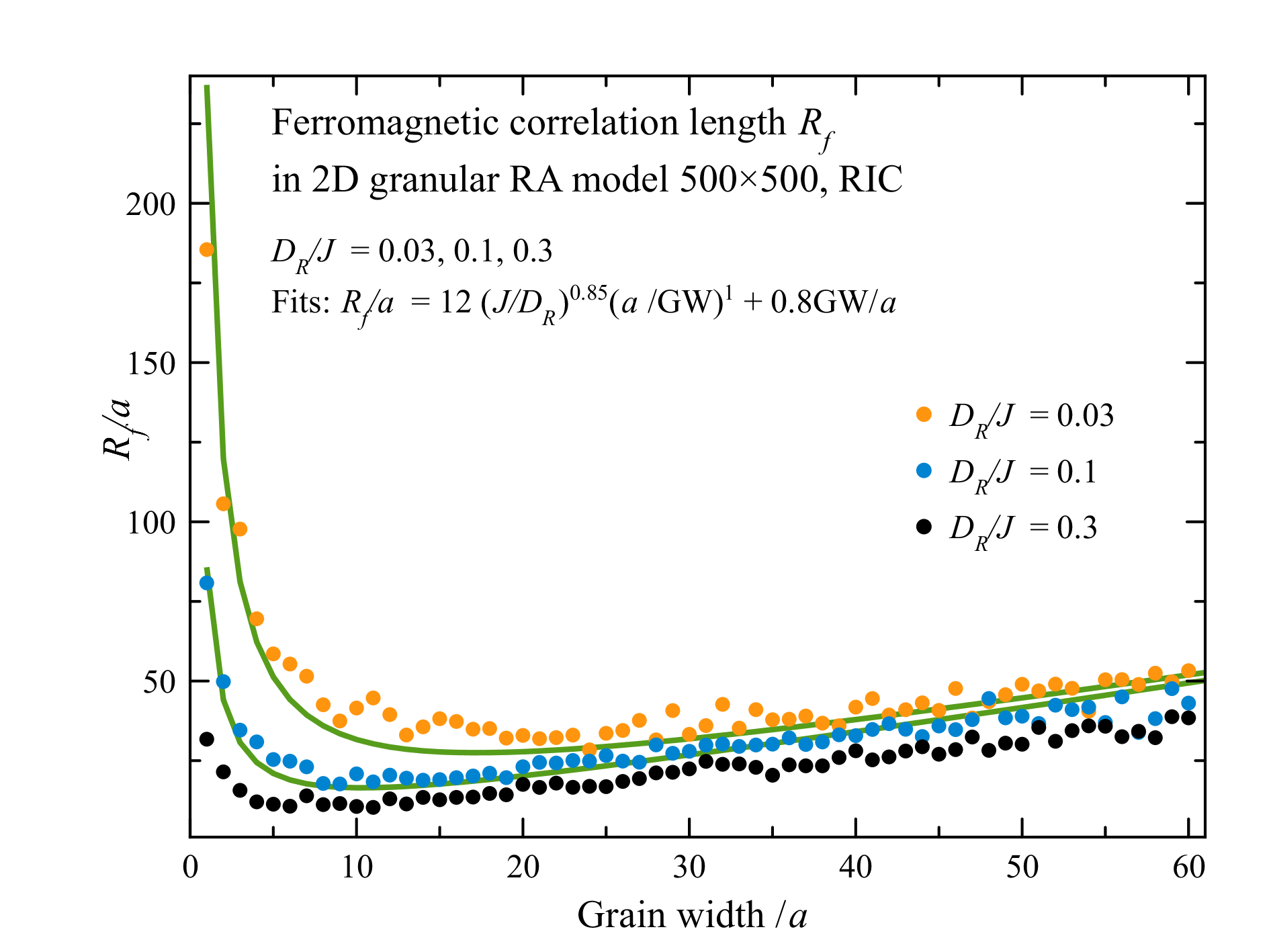}\
\caption{Dependence of the ferromagnetic correlation length on the size of the grain illustrating the crossover from the collective magnetic behavior to the independent magnetic ordering within individual grains in a two-dimensional granular ferromagnet.}
\label{Rf_vs_GrainSize} 
\end{figure}

The requirement that $R_f$ cannot go below $a$ establishes the range of validity of the above formulas on the size of the grain:
\begin{equation}
\frac{R_a}{a} < k_2\left(\frac{J}{D_R}\right) = \left(\frac{R_f}{a}\right)_{R_a = a}
\end{equation} 
in two dimensions, and
\begin{equation}
\frac{R_a}{a} < k_3^{1/3}\left(\frac{J}{D_R}\right)^{2/3} = \left(\frac{R_f}{a}\right)^{1/3}_{R_a = a}
\end{equation} 
in three dimensions. As $R_f$ decreases on increasing $R_a$, this condition breaks at  $R_f(R_a) \sim R_a$. At that point, the system becomes equivalent to an array of densely packed, weakly interacting, single-domain ferromagnetic particles of size $R_a$ with $R_f = R_a$. The crossover of the granular ferromagnet from the collective behavior extending over many grains to the independent magnetic ordering within individual grains is illustrated in Fig.\ \ref{Rf_vs_GrainSize}.

\section{Microwave absorption}
\label{absorption}

In this section, we will show how the scaling arguments explored in the previous section help understand microwave absorption by RA magnets. In numerical work, we solve the discrete model with atomic spins. All computations use the fluctuation-dissipation theorem as explained in Ref.\ \onlinecite{GC-PRB2022}. 

\subsection{Peak frequency}
\label{peak}

In Ref.\  \onlinecite{GC-PRB2021} we argued that the absorption of microwaves by the RA magnet is dominated by excitation modes that can be interpreted as the FMR of regions of size $R_f$. In a crystalline ferromagnet with the anisotropy constant $D$ this would correspond to the FMR frequency $D/\hbar$ in a zero field if dipole-dipole interactions were neglected. Along this line, the absorption peak in an amorphous ferromagnet must be determined by statistical fluctuations of the anisotropy inside ferromagnetically correlated regions. The corresponding frequency is given by $(D_R/\hbar)(a/R_f)^{d/2}$, which was confirmed by numerical work for $d = 1,2,3$. 

Generalization for the peak frequency in a granular ferromagnet is straightforward:
\begin{equation}
\omega_{max} = D_R\left(\frac{R_a}{R_f}\right)^{d/2}.
\end{equation}
Substituting here Eq.\ (\ref{Rf-Ra}), we obtain
\begin{equation}
\omega_{max} = \frac{D_R}{k_d^{d/2}}\left(\frac{D_R}{J}\right)^{d/(4-d)}\left(\frac{R_a}{a}\right)^{2d/(4-d)},
\label{omega-max-d}
\end{equation}
which gives
\begin{equation}
\omega_{max} = \frac{D_R}{k_2}\left(\frac{D_R}{J}\right)\left(\frac{R_a}{a}\right)^2
\label{omega-max-2D}
\end{equation} 
in two dimensions and
\begin{equation}
\omega_{max} = \frac{D_R}{k_3^{3/2}}\left(\frac{D_R}{J}\right)^3\left(\frac{R_a}{a}\right)^6
\end{equation} 
in three dimensions. 

Notice a very strong dependence on $D_R/J$ and $R_a$ in three dimensions. For physical values $D_R$ that are smaller than $J$ by orders of magnitude, one should choose sufficiently large grains if the absorption in the GHz range is desired. As the grain size $R_a$ approaches the domain wall thickness $a \sqrt{J/D_R}$, the FM correlation length $R_f$ for all $d$ approaches $R_a$, and, as follows from the above equations, the peak frequency in both two and three dimensions tends to the FMR frequency $D_R/\hbar$ of the grain. 

\begin{figure}[h]
\centering{}\includegraphics[width=9cm]{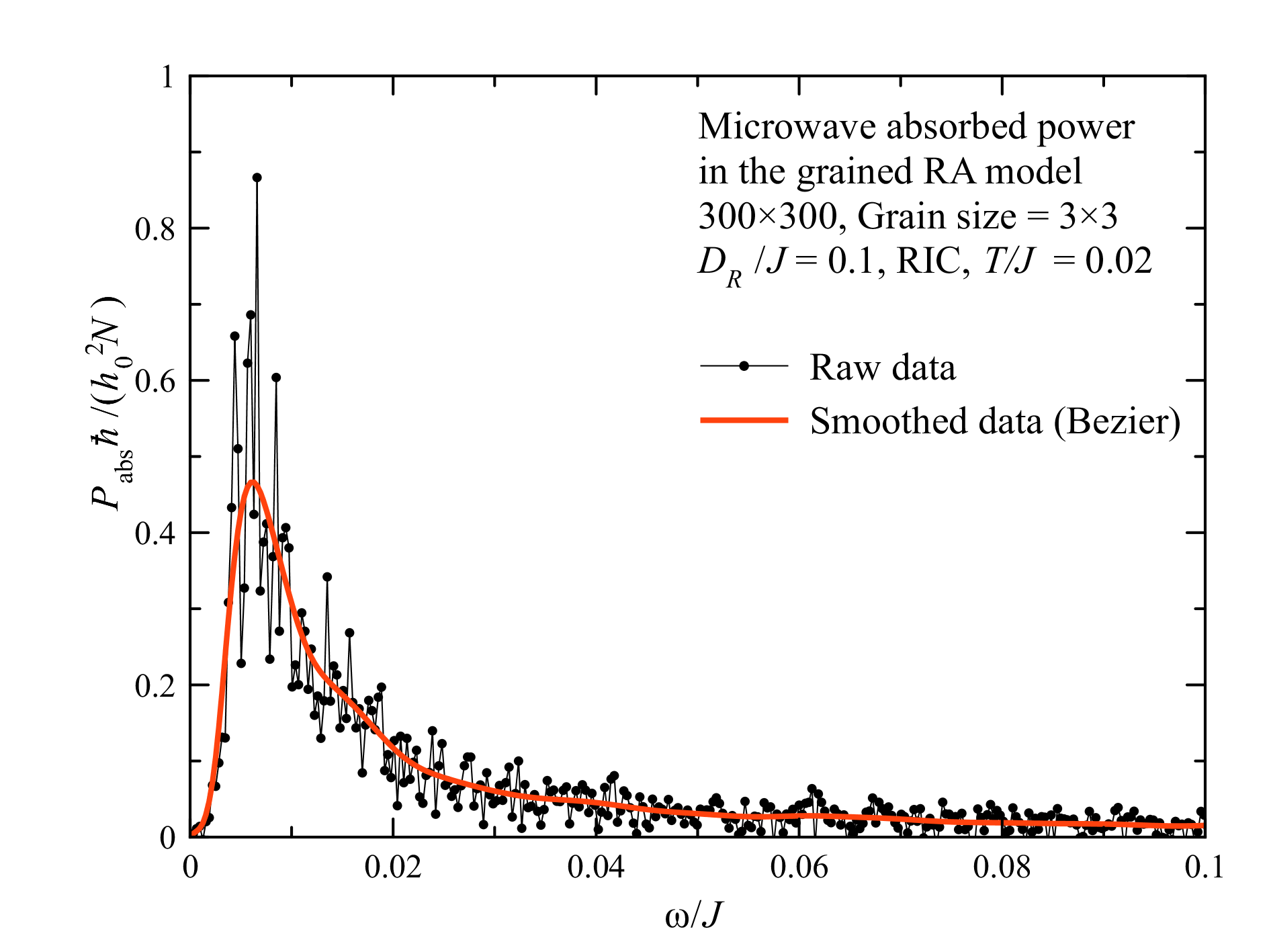}\
\caption{Frequency dependence of the absorbed microwave power for a 2D system composed of $3\times 3$ grains with $D_R/J = 0.1$. B\'{e}zier functions have been used to draw a smooth curve.}
\label{Bezier} 
\end{figure}
Frequency dependence of the absorbed power for a 2D system composed of  $3\times 3$ grains with $D_R/J = 0.1$ is shown in Fig.\ \ref{Bezier}. The numerical method has been described in detail in Ref.\  \onlinecite{GC-PRB2022}. For small $D_R$ local modes \cite{GC-PRB2023L} are sparse and the statistics is not great even for system sizes as large as $300 \times 300$. We used B\'{e}zier functions \cite{Bezier} to draw a smooth curve.

\begin{figure}[h]
\centering{}\includegraphics[width=9cm]{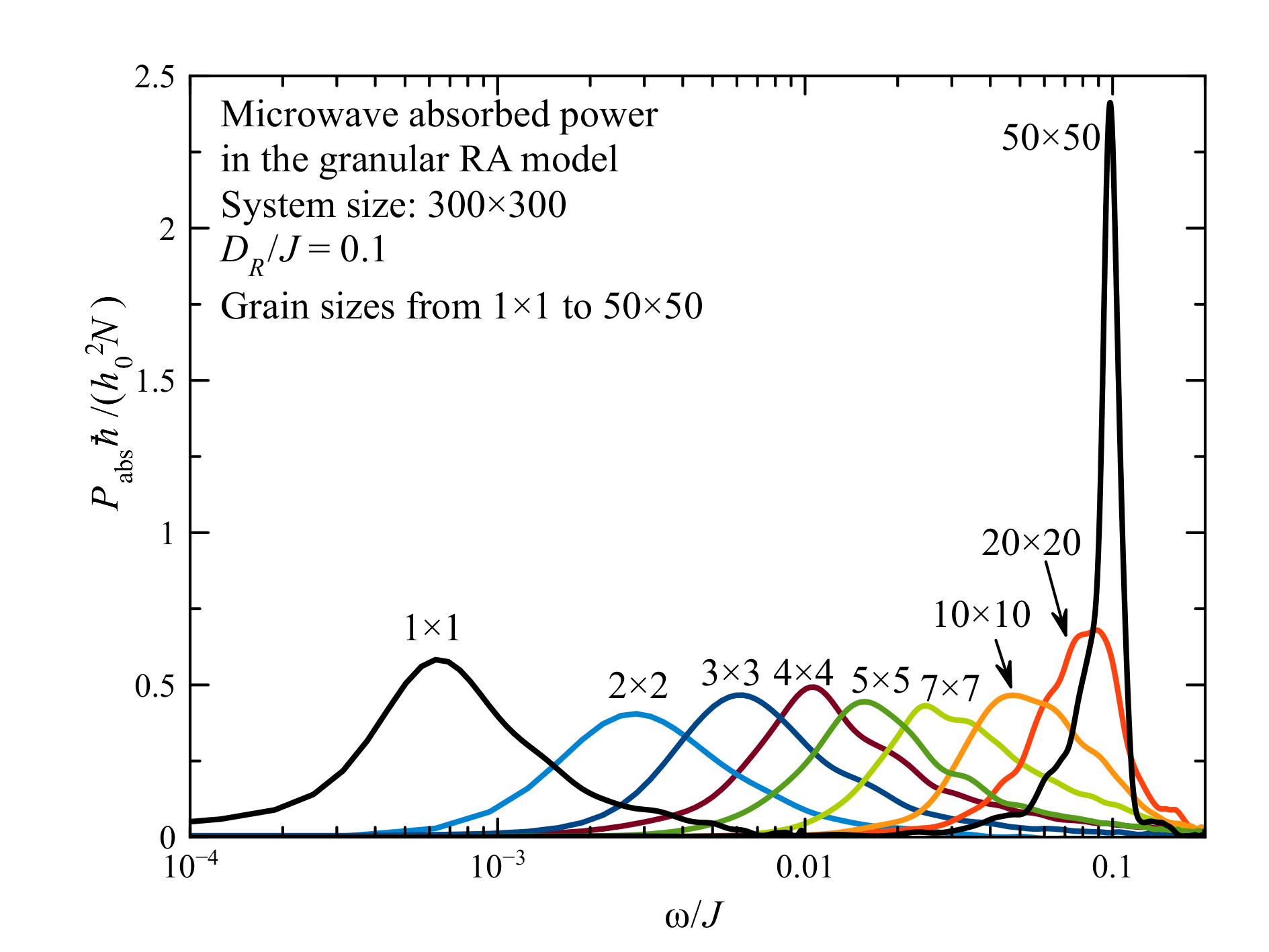}\
\caption{Frequency dependence of the absorbed microwave power for different grain sizes computed on a $300 \times 300$ spin lattice at $D_R/J = 0.1$. B\'{e}zier functions for raw data have been used.}
\label{P-omega} 
\end{figure}
\begin{figure}[h]
\centering{}\includegraphics[width=9cm]{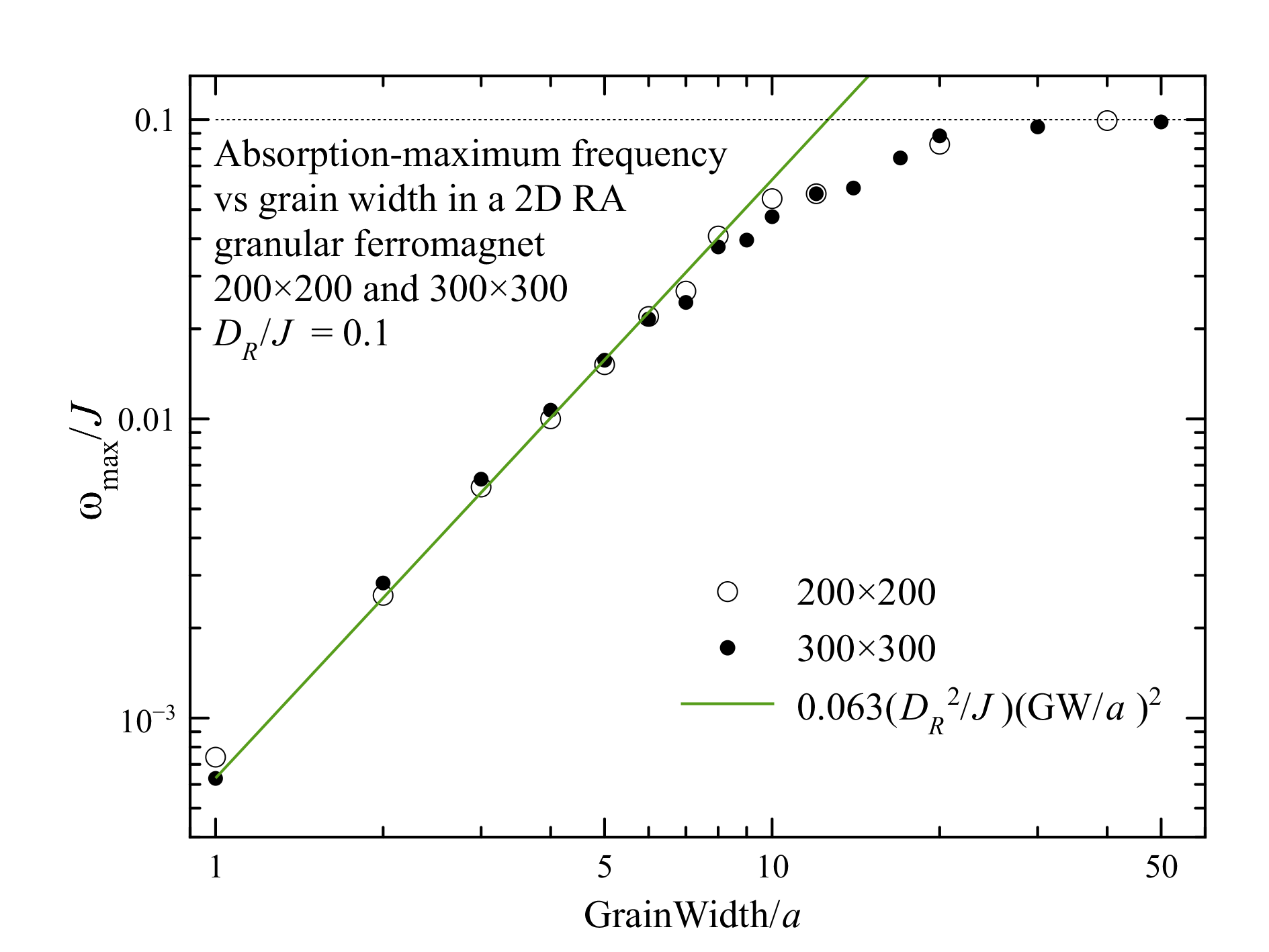}\
\caption{Dependence of the peak frequency in the absorption of microwaves by a granular ferromagnet on the size of the grain.}
\label{omega-max} 
\end{figure}

Fig.\ \ref{P-omega} shows the frequency dependence of the absorbed power by a granular ferromagnet with different grain sizes. For small grains, the dependence of the peak frequency on the size of the grain, shown in Fig.\ \ref{omega-max}, follows the theoretical formula (\ref{omega-max-2D}). In agreement with the theory, it switches to $\omega_{max} = D_R/\hbar$ for large grains. The peak in Fig.\ \ref{P-omega} becomes narrow, tending to the delta-function at $R_a \rightarrow \infty$ in the absence of dissipation. 

\subsection{Power absorption}
\label{power}

\begin{figure}[h]
\centering{}\includegraphics[width=9cm]{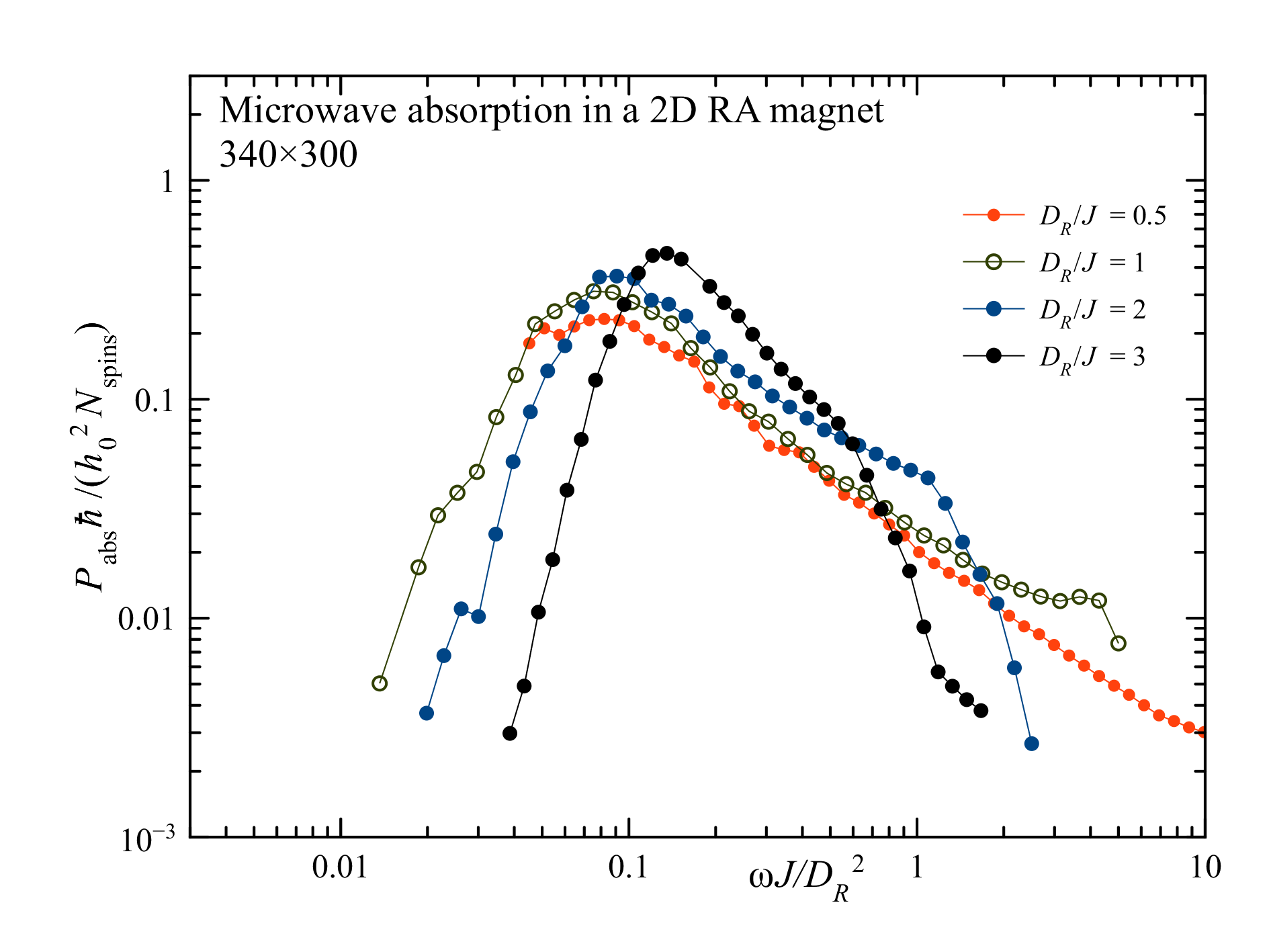}\
\caption{Scaled dependence of the absorbed power on the combination of frequency and RA strength in two dimensions. The unscaled data are taken from Fig. 4 of Ref.\ \onlinecite{GC-PRB2021}.}
\label{P-scaled} 
\end{figure}

In Ref.\  \onlinecite{GC-PRB2021} we have shown that 
\begin{equation}
P \sim h_0^2N \frac{D_R^2}{J^2} \left(\frac{J}{\omega}\right)^{(4-d)/2},
\label{power-21}
\end{equation}
(with $h_0$ being the amplitude of the microwave field and $N$ being the total number of spins) provides a good fit for the microwave absorption at $\omega > \omega_{max}$ by a $d$-dimensional RA ferromagnet with $N$ spins and atomic-scale disorder. Cases of $d=1,2,3$ were studied numerically. Eq.\ (\ref{power-21}) also explains the numerical finding that the peak power at $\omega = \omega_{max}$ is weakly dependent on $D_R$ in all dimensions. Indeed, for the atomic disorder with $D_R \ll J$, substituting  Eq. (\ref{omega-max-d}) at $R_a = a$ into Eq.\ (\ref{power-21}), gives $P_{max} \sim h_0^2N$, which is independent of $D_R$. The proportionality of the power to $D_R^2/\omega$ in two dimensions and weak dependence of the peak power on $D_R$ are illustrated by Fig.\ \ref{P-scaled}.

We shall now try to come up with the dependence of the power at $\omega > \omega_{max}$ on the size of the grain in a granular ferromagnet by writing it in the form
\begin{equation}
P \sim \frac{h_0^2}{J}\left(N' \frac{{D'}_R^2}{J'} \right) \left(\frac{J}{\omega}\right)^{(4-d)/2}, 
\end{equation}
with the rescaled exchange and anisotropy constants given by Eq.\ (\ref{scaling}) and
\begin{equation}
N' = N\left(\frac{a}{R_a}\right)^{d}
\label{grain-number}
\end{equation}
for the number of grains. The reason we are rescaling that particular combination $N' {{D'}_R^2}/{J'} $ is that it determines the power in the second order of the perturbation theory on RA. The rescaling yields
\begin{equation}
\frac{{D'_R}^2}{J'} N' = \frac{{D_R}^2}{J} N \left(\frac{R_a}{a}\right)^2,
\label{scaling-prime}
\end{equation}
making the power absorbed by the grains given by
\begin{equation}
P_g \sim h_0^2N \frac{D_R^2}{J^2} \left(\frac{R_a}{a}\right)^2\left(\frac{J}{\omega}\right)^{(4-d)/2}.
\label{P-g}
\end{equation}
Substituting here Eq.\ (\ref{omega-max-d}) for $\omega$, we obtain for the peak power
\begin{equation}
(P_g)_{max} \sim h_0^2N\left(\frac{a}{R_a}\right)^{d-2}.
\label{peak-power}
\end{equation}
In two dimensions it does not depend on $R_a$. In three dimensions it scales as $a/R_a$. We speculate that the weak dependence of the peak on the grain size at $R_a < R_f$, seen in two dimensions in Fig.\ \ref{P-omega}, may be due to energy barriers and topological defects unaccounted for in the theoretical model.

We can double-check these formulas by computing the integral power (IP), $\int d\omega P(\omega)$, for which we have independent theory confirmed by numerics, see Ref.\  \onlinecite{CG-PRB2023IP}. The bulk of the IP comes from $\omega > \omega_{max}$. For $d = 2$ we obtain from Eq.\ (\ref{P-g})
\begin{eqnarray}
({\rm IP})_2 & \sim & h_0^2N \frac{D_R^2}{J^2} \left(\frac{R_a}{a}\right)^2 J \int^J_{\omega_{max}} \frac{d\omega}{\omega} \nonumber \\
& = & h_0^2N \frac{D_R^2}{J} \left(\frac{R_a}{a}\right)^2 \ln(J/\omega_{max}) \nonumber \\
& = & h_0^2N \frac{D_R^2}{J} \left(\frac{R_a}{a}\right)^2 \ln\left[C\left(\frac{J}{D_R}\right)\left(\frac{a}{R_a}\right)\right]
\end{eqnarray}
with $C$ being the integration constant of order unity. This is exactly what we had for the IP in two dimensions \cite{CG-PRB2023IP}. For $d = 3$ we get 
\begin{eqnarray}
({\rm IP})_3 & \sim & h_0^2N \frac{D_R^2}{J^2} \left(\frac{R_a}{a}\right)^2 \sqrt{J} \int^J_{\omega_{max}} \frac{d\omega}{\sqrt{\omega} } \nonumber \\
& \sim & h_0^2N \frac{D_R^2}{J} \left(\frac{R_a}{a}\right)^2, 
\end{eqnarray}
which is again exactly what we had for the IP in three dimensions \cite{CG-PRB2023IP}. 

Thus, with good confidence, the scaling of the absorbed power with the frequency and the grain size for a granular ferromagnet in $d$ dimensions at $\omega > \omega_{max}$ is given by Eq.\ (\ref{P-g}). 

\section{Discussion}
\label{discussion}

The RA model has been used in the past to describe amorphous and sintered ferromagnets. Regardless of the strength $D_R$ of the RA that produces the static random potential for the spins locally ordered by the ferromagnetic exchange $J$, the model is non-perturbative  on $D_R/J$. In amorphous ferromagnets, this parameter is very small due to the relativistic smallness of the magnetic anisotropy. This always hindered numerical work on real systems because it required a linear size of the system that is large compared to the ferromagnetic correlation length which scales as $J/D_R$ in two dimensions and $(J/D_R)^2$ in three dimensions. Consequently, even a very large (by physical standards) value of $D_R/J$, such as, e.g., $10^{-3}$, requires modeling of systems that consist of the prohibitively large number of spins. 

Here we have demonstrated that due to the existence of the scaling specific to static randomness, the RA model of an amorphous ferromagnet with the atomic-scale disorder and $D_R \ll J$ can be mapped onto the model with $D_R \sim J$. This makes static magnetic properties of the magnet, such as magnetization, invariant with respect to $D_R/J$ up to a critical large value of that ratio. Consequently, one can make conclusions about the behavior of a large system with realistic small values of the magnetic anisotropy by modeling much smaller systems with $D_R \sim J$. 
It must somehow be related to the fractal structure of the random walk which is behind the Imry-Ma argument, that is the invariance of the random walk with respect to the rescaling of its trajectory to a map with a different scale. 

The scaling concept is also extremely helpful in application to granular ferromagnets for which the effective anisotropy constant $D'_R$ scales as the size of the grain squared in two dimensions (thin film) and as the size of the grain cubed in three dimensions, while the exchange $J'$ equals the original exchange constant in two dimensions and scales linearly on the grain size in three dimensions. This, in principle, permits any values of the ratio $J'/D'_R$ in experiments with granular ferromagnets depending on the grain size. The scaling breaks down at the critical grain size of the order of the domain wall width in a conventional ferromagnet, $\delta = a\sqrt{J/D_R}$.  Above such a grain size the behavior of the granular magnet changes qualitatively. The ferromagnetic correlation length, which was going down on increasing the grain size below the critical size, begins to grow with the grain size above the critical size. As a result, the static properties of the granular magnet switch from soft to hard magnetic behavior.

Using the scaling arguments, we also studied the absorption of microwave power by the granular ferromagnet. The peak frequency has been expressed in terms of the magnetic anisotropy and the size of the grain and was tested in a numerical experiment. The dependence of the power on frequency, magnetic anisotropy, and the size of the grain have been obtained analytically and numerically, with good agreement between the two methods.  Our results apply to insulating amorphous materials or materials made of coated metallic ferromagnetic grains of size which is small compared to the skin depth, which would typically be in the submicron range for the microwaves.  

The breakdown of the scaling at the critical value of the grain size, besides its effect on the static properties, has a dramatic effect on microwave absorption. On increasing the grain size above the critical value, the absorption peak narrows and increases in height. It makes the grain size of the order of the domain wall width (in the ferromagnetic material of the grain) optimal for achieving high absorption and the broad band at the same time. These predictions can be tested on recently synthesized numerous insulating amorphous/nanocrystalline magnetic materials \cite{AlAzzavi-2016,Lu-JNCS2023}.

Scaling concepts developed in this paper can be used for other systems with quenched randomness. They should facilitate numerical studies of disordered materials by extending conclusions made from the modeling of small systems with a strong static disorder to large systems with a weak disorder. In application to random magnets, they should assist material scientists in manufacturing materials with desired magnetic and microwave properties.  

\section*{ACKNOWLEDGEMENT}

This work has been supported by Grant No. FA9550-20-1-0299 funded by the Air Force Office of Scientific Research.


\begin{thebibliography}{10}

\bibitem{RA-book} E. M. Chudnovsky, Random Anisotropy in Amorphous
Alloys, Chapter 3 in the Book: \textit{Magnetism of Amorphous Metals
and Alloys}, edited by J. A. Fernandez-Baca and W.-Y. Ching, pages
143-174 (World Scientific, Singapore, 1995).

\bibitem{CT-book} E. M. Chudnovsky and J. Tejada, \textit{Lectures
on Magnetism} (Rinton Press, Princeton, New Jersey, 2006).

\bibitem{Marin-MMM2020}
P. Marin and A. Hernando, Applications of amorphous and nanocrystalline magnetic materials, Journal of Magnetism and Magnetic Materials {\bf 215-216}, 729-734 (2020).

\bibitem{GC-JPhys2022}
D. A. Garanin and E. M. Chudnovsky, Random anisotropy magnet at finite temperature, Journal of Physics: Condensed Matter {\bf 34}, 285801-(15) (2022).

\bibitem{bubbles}
R. Seshadri and R.M. Westervelt, Statistical mechanics of magnetic bubble arrays. I. Topology and thermalization, Physical Review B {\bf 46},  5142-5149 (1992); ibid. II. Observations of two-dimensional melting, {\bf 46}, 5150-5161 (1992). 

\bibitem{Efetov-77}
K. B. Efetov and A. I. Larkin, Charge-density wave in a random potential, Soviet Physics JETP {\bf 45}, 1236-1241 (1977). 

\bibitem{Okamoto-2015}
J.-i. Okamoto, C. J. Arguello, E. P. Rosenthal, A. N. Pasupathy, and A J. Millis, Experimental evidence for a Bragg glass density wave phase
in a transition-metal dichalcogenide, Physical Review Letters {\bf 114}, 026802-(5) (2015). 

\bibitem{Blatter-RMP1994} 
See, e.g., G. Blatter, M. V. Feigel'man, V. B. Geshkenbein, A.I. Larkin, and V. M. Vinokur, Vortices in high-temperature superconductors, Review of Modern Physics {\bf 66}, 1125-1388 (1994), and references therein.

\bibitem{EC-PRB1989}
E. M. Chudnovsky, Hexatic vortex glass in disordered superconductors, Physical Review B {\bf 40}, 11355-11357 (1989).

\bibitem{LC}
T. Bellini, N. A. Clark, V. Degiorgio, F. Mantegazza, and G. Natale, Light-scattering measurement of the nematic correlation length in a liquid crystal with quenched disorder, Physical Review E {\bf 57}, 2996-3006 (1998). 

\bibitem{EC-PRB1986}
E. M. Chudnovsky, Structure of a solid film on an imperfect surface, Physical Review B {\bf 33}, 245-250 (1986).

\bibitem{Volovik-JLTP2008}
G. E. Volovik, On Larkin-Imry-Ma state in $^4$He-A in aerogel, Journal of Low Temperature Physics {\bf 150}, 453-463 (2008).

\bibitem{Li-Nature2013}
J. I. Li, J. Pollanen, A. M. Zimmerman, C. A. Collett, W. J. Gannon, and W.P. Halperin, The superfluid glass phase of $^3$He-A, Nature Physics {\bf 9}, 775-779 (2013).

\bibitem{Volovik-JETPlett2018}
G. E. Volovik, Topology of a $^3$He-A film on a corrugated graphene substrate, JETP Letters {\bf 2107}, 115-118 (2018).

\bibitem{IM} 
Y. Imry and S.-k. Ma, Random-field instability of the ordered state of continuous symmetry, Physical Review Letters \textbf{35}, 1399-1401 (1975).

\bibitem{Harris-PRL1973}
R. Harris, M. Plischke, and M. J. Zuckermann, New model for amorphous magnetism, Physical Review Letters {\bf 31}, 160-162 (1973).

\bibitem{CSS-1986} 
E. M. Chudnovsky, W. M. Saslow, and R. A. Serota, Ordering in ferromagnets with random anisotropy, Physical Review B \textbf{33}, 251-261 (1986).

\bibitem{PCG-2015} 
T. C. Proctor, E. M. Chudnovsky, and D. A. Garanin, Scaling of coercivity in a 3d random anisotropy model, Journal of Magnetism and Magnetic Materials, \textbf{384}, 181-185 (2015).

\bibitem{Cardy-PRB1982} 
J. L. Cardy and S. Ostlund, Random symmetry-breaking fields and the  XY model, Physical Review B {\bf 25}, 6899-6909 (1982).

\bibitem{Villain-ZPB1984} 
J. Villain and J. F. Fernandez, Harmonic system in a random field, Zeitschrift f\"ur Physik B - Condensed Matter {\bf 54}, 139-150 (1984).

\bibitem{Nattermann}
T. Nattermann, Scaling approach to pinning: Charge density waves and giant flux creep in superconductors, Physical Review Letters {\bf 64}, 2454-2457 (1990).

\bibitem{Kierfield}
J. Kierfield, T. Nattermann, and T. Hwa, Topological order in the vortex-glass phase of high-temperature superconductors, Physical Review B {\bf 55}, 626-629 (1997).

\bibitem{Korshunov-PRB1993} 
S. E. Korshunov, Replica symmetry breaking in vortex glasses, Physical Review B {\bf 48}, 3969-3975 (1993).

\bibitem{Giamarchi-95}
T. Giamarchi and P. Le Doussal, Elastic theory of flux lattices in the presence of weak disorder, Physical Review B {\bf 52}, 1242-1270 (1995).

\bibitem{LeDoussal-PRL07}
A. A. Middleton, P. Le Doussal, and K. J. Wiese, Measuring Functional Renormalization Group Fixed-Point Functions for Pinned Manifolds, Physical Review Letters {\bf 98}, 155701-(4) (2007).

\bibitem{Bogner}
S. Bogner, T. Emig, A. Taha, and C. Zeng, Test of replica theory: Thermodynamics of two-dimensional model systems with quenched disorder, Physical Review B {\bf 69}, 104420-(15) (2004).

\bibitem{Orland-EPL}
H. Orland and Y. Shapir, A Disorder-Dependent Variational Method Without Replicas: Application to the Random Phase Sine-Gordon Model, Europhysics Letters {\bf 30}, 203 (1995).

\bibitem{Garel-PRB}
T. Garel, G. Lori, and H. Orland, Variational study of the random-field XY model, Physical Review B {\bf 53}, R2941-R2944 (1996).

\bibitem{Gingras-Huse-PRB1996} 
M. J. P. Gingras and D. A. Huse, Topological defects in the random-field XY model and the pinned vortex lattice to vortex glass transition in type-II superconductors, Physical Review B {\bf 53}, 15193-15200 (1996).

\bibitem{Zeng} 
C. Zeng, A. A. Middleton, and Y. Shapir, Ground-State Roughness of the Disordered Substrate and Flux Lines in d=2, Physical Review Letters {\bf 77}, 3204-3207 (1996).

\bibitem{Rieger}
H. Rieger and U. Blasum, Ground-state properties of solid-on-solid models with disordered substrates, Physical Review B {\bf 55}, R7394-R7397 (1997).

\bibitem{Itakura-05}
M. Itakura and C. Arakawa, Quasi Long Range Ordered Ground-State of the Random Field XY Mode, Progress in Theoretical Physics Supplements {\bf 157}, 136-138 (2005).

\bibitem{Perret-PRL2012}
A. Perret, Z. Ristivojevic, P. Le Doussal, G. Schehr, and K. J. Wiese, Super-rough glassy phase of the random field XY Model in two dimensions, Physical Review Letters {\bf 109}, 157205-(5) (2012).

\bibitem{Fisch-1998} 
R. Fisch, Quasi-long-range order in random-anisotropy Heisenberg models, Physical Review B {\bf 58}, 5684-5691 (1998).

\bibitem{Fisch-2000}
R. Fisch, Random field and random anisotropy effects in defect-free three-dimensional XY models, Physical Review B {\bf 62}, 361-366 (2000).

\bibitem{GCP-PRB2013}
D. A. Garanin, E. M. Chudnovsky, and T. Proctor, Random field $xy$ model in three dimensions, Physical Review B {\bf 88}, 224418-(21) (2013).

\bibitem{PGC-PRL} 
T. C. Proctor, D. A. Garanin, and E. M. Chudnovsky,
Random fields, Topology, and Imry-Ma argument, Physical Review Letters
\textbf{112}, 097201-(4) (2014).

\bibitem{CG-PRL} E. M. Chudnovsky and D. A. Garanin, Topological
order generated by a random field in a 2D exchange model, Physical
Review Letters \textbf{121}, 017201-(4) (2018).

\bibitem{Serota-1986}
R.A. Serota and P.A. Lee, Continuous-symmetry ferromagnet with weal magnetic anisotropy: A numerical study, Physical Review B {\bf 34}, 1806-1810 (1986).

\bibitem{Dieny-PRB1990} 
B. Dieny and B. Barbara, XY model with weak random anisotropy in a symmetry-breaking magnetic field, Physical Review B {\bf 41}, 11549-11556 (1990).

\bibitem{DC-PRB1991} 
R. Dickman and E. M. Chudnovsky, XY chain with random anisotropy: Magnetization law, susceptibility, and correlation functions at T=0, Physical Review B {\bf 44}, 4397-4405 (1991).

\bibitem{Saslow2018} W. M. Saslow and C. Sun, Longitudinal resonance
for thin film ferromagnets with random anisotropy, Physical Review
B \textbf{98}, 214415-(6) (2018).

\bibitem{Monod} P. Monod and Y. Berthier, Zero field electron spin
resonance of Mn in the spin glass state, Journal of Magnetism and
Magnetic Materials \textbf{15-18},149-150 (1980).

\bibitem{Prejean} J. J. Prejean, M. Joliclerc, and P. Monod, Hysteresis
in CuMn: The effect of spin-orbit scattering on the anisotropy in
the spin glass state, Journal de Physique (Paris) \textbf{41}, 427-435
(1980).

\bibitem{Alloul1980} H. Alloul and F. Hippert, Macroscopic magnetic
anisotropy in spin glasses: transverse susceptibility and zero field
NMR enhancement, Journal de Physique Lettres \textbf{41}, L201-204
(1980).

\bibitem{Schultz} S. Schultz, E .M. Gulliksen, D. R. Fredkin, and
M.Tovar, Simultaneous ESR and magnetization measurements characterizing
the spin-glass state, Physical Review Letters \textbf{45}, 1508-1512
(1980).

\bibitem{Gullikson} E. M. Gullikson, D. R. Fredkin, and S. Schultz,
Experimental demonstration of the existence and subsequent breakdown
of triad dynamics in the spin-glass CuMn, Physical Review Letters
\textbf{50}, 537-540 (1983).

\bibitem{Fert} A. Fert and P. M. Levy, Role of anisotropic exchange
interactions in determining the properties of spin-glasses, Physical
Review Letters \textbf{44},1538-1541 (1980).

\bibitem{Levy} P. M. Levy and A. Fert, Anisotropy induced by nonmagnetic
impurities in CuMn spin-glass alloys, Physical Review B \textbf{23},
4667 (1981).

\bibitem{Henley1982} C. L. Henley, H. Sompolinsky, and B. I. Halperin,
Spin-resonance frequencies in spin-glasses with random anisotropies,
Physical Review B \textbf{25}, 5849-5855, (1982).

\bibitem{HS-1977} B. I. Halperin and W. M. Saslow, Hydrodynamic theory
of spin waves in spin glasses and other systems with noncollinear
spin orientations, Physical Review B \textbf{16}, 2154-2162 (1977).

\bibitem{Saslow1982} W. M. Saslow, Anisotropy-triad dynamics, Physical
Review Letters \textbf{48}, 505-508 (1982).

\bibitem{Bruinsma1986}
R. Bruinsma and S. N. Coppersmith, Anderson localization and breakdown of hydrodynamics in random ferromagnets, Physical Review B {\bf 33}, 6541-6544(R) (1986).

\bibitem{Serota1988} 
R. A. Serota, Spin-wave localization in ferromagnets with weak random anisotropy, Physical Review B {\bf 37}, 9901-9903(R) (1988).

\bibitem{Ma-PRB1986}
M. Ma, B. I. Halperin, and P.A. Lee, Strongly disordered superfluids: Quantum fluctuations and critical behavior, Physical Review B {\bf 34}, 3136-3143 (1986).

\bibitem{Zhang-PRB1993}
L. Zhang, Disordered boson systems: A perturbative study, Physical Review B {\bf 47}, 14364-14373 (1993).

\bibitem{Alvarez-PRL2013}
J. P. \'{A}lvarez Z\'{u}\~{n}iga and N. Laflorencie, Bose-glass transition and spin-wave localization for 2D bosons in a random potential, Physical Review Letters {\bf 111}, 160403-(5) (2013).

\bibitem{Yu-AnnPhys2013}
X. Yu and M. M\"{u}ller, Localization of disordered bosons and magnets in random fields, Annals of Physics {\bf 337}, 55-93 (2013).

\bibitem{Nowak2015}
M. Evers, C. A. M\"{u}ller, and U. Nowak, Spin-wave localization in disordered magnets, Physical Review B {\bf 92}, 014411 (2015).

\bibitem{Amaral-1993}
V. S. Amaral, B. Barbara, J. B. Sousa, and J. Filippi, Spin-wave localization in random anisotropy systems: Amorphous (Dy$_x$Gd$_{1-x}$)Ni, European Physics Letters {\bf 22} 139-144 (1993).

\bibitem{Suran1-1997}
S. Suran and E. Boumaiz, Longitudinal resonance in ferromagnets with random anisotropy: A formal experimental demonstration, Journal of Applied Physics \textbf{81}, 4060 (1997).

\bibitem{Suran2-1997}
G. Suran, Z, Frait, and E. Boumaz, Direct observation of the longitudinal resonance mode in ferromagnets with random anisotropy, Physical Review B \textbf{55}, 11076-11079 (1997).

\bibitem{Suran-1998} 
S. Suran and E. Boumaiz, Longitudinal-transverse resonance and localization related to the random anisotropy in a-CoTbZr films, Journal of Applied Physics \textbf{83}, 6679 (1998).

\bibitem{McMichael-PRL2003}
R. D. McMichael, D. J. Twisselmann, and A. Kunz, Localized ferromagnetic resonance in inhomogeneous thin film, Physical Review Letters \textbf{90}, 227601-(4) (2003).

\bibitem{Loubens-PRL2007} 
G. de Loubens, V. V. Naletov, O. Klein, J. Ben Youssef, F. Boust, and N. Vukadinovic, Magnetic resonance studies of the fundamental spin-wave modes in individual submicron Cu/NiFe/Cu
perpendicularly magnetized disks, Physical Review Letters \textbf{98}, 127601-(4) (2007).

\bibitem{Du-PRB2014}
C. Du, R. Adur, H. Wang, S. A. Manuilov, F. Yang, D. V. Pelekhov, and P. C. Hammel, Experimental and numerical understanding of localized spin wave mode behavior in broadly tunable
spatially complex magnetic configurations, Physical Review B \textbf{90}, 214428-(10) (2014).

\bibitem{GC-PRB2021} 
D. A. Garanin and E. M. Chudnovsky, Absorption
of microwaves by random-anisotropy magnets, Physical Review B \textbf{103},
214414-(11) (2021).

\bibitem{GC-PRB2022}
D. A. Garanin and E. M. Chudnovsky, Nonlinear and thermal effects in the absorption of microwaves by random anisotropy magnets, Physical Review B {\bf 105}, 064402-(8) (2022). 

\bibitem{GC-PRB2023L}
D. A. Garanin and E. M. Chudnovsky, Localized spin-wave modes and microwave absorption in random-anisotropy ferromagnets, Physical Review B {\bf 107}, 134411-(13) (2023).

\bibitem{CG-PRB2023IP}
E. M. Chudnovsky and D. A. Garanin, Integral absorption of microwave power by random-anisotropy magnets, Physical Review B {\bf 107}, 224413-(9) (2023).

\bibitem{Bezier}
H. Prautzsch, W. Boehm, and M. Paluszny, {\it B\'{e}zier and B-Spline Techniques} (Springer Berlin, Heidelberg, 2002).

\bibitem{AlAzzavi-2016}
H. S. M. Al'Azzavi, A. B. Granovskii, Yu. E. Kalinin, V. A. Makagonov, A. V. Sitnikov, and O. S. Tarasova, Influence of oxidized interlayers on magnetic properties of multilayer films
based on amorphous ferromagnet-dielectric nanocomposites, Physics of Solid State {\bf 58}, 938-945 (2016).

\bibitem{Lu-JNCS2023}
S. Lu, M. Wang, and Z. Zhao, Recent advances and future developments in Fe-based amorphous soft magnetic composites, Journal of Non-Crystalline Solids {\bf 616}, 122440-(16) (2023).

\end{thebibliography}
\end{document}